\documentclass[11pt]{article}
\usepackage{amssymb}
\usepackage{amscd}
\usepackage{amsmath}

\catcode`@=11
\renewcommand{\section}{\@startsection{section}{1}{0pt}{\medskipamount}
{\medskipamount}{\large\bf}}
\catcode`@=12

\def\a{\alpha}
\def\b{\beta}

\def\de{\delta}
\def\eps{\epsilon}

\def\l{\lambda}

\def\s{\sigma}
\def\t{\tau}
\def\x{\xi}

\def\j{\psi}

\newcommand{\R}{\mathbb R}
\newcommand{\Z}{\mathbb Z}

\def\e{\textrm{e}}
\def\i{\textrm{i}}
\def\N2{$N{=}2$}
\def\NN4{$N{=}4$}
\def\pa{\mbox{$\partial$}}

\def\sfrac#1#2{{\textstyle\frac{#1}{#2}}}
\def\shalf{\sfrac{1}{2}}

\def\ad{{\dot{\alpha}}}

\def\zd{{\dot{0}}}
\def\od{{\dot{1}}}
\def\add{{\ddot{\alpha}}}
\def\bdd{{\ddot{\beta}}}

\def\zdd{{\ddot{0}}}
\def\odd{{\ddot{1}}}

\def\pinv{\sfrac{1}{p^{0\zd}}}
\def\kinv{\sfrac{1}{k^{0\zd}}}

\begin{document}
\begin{titlepage}
\setcounter{page}{0}
\begin{flushright}
hep-th/0204073\\
ITP--UH--06/02\\
YITP-SB-02-16
\end{flushright}

\vskip 2.0cm

\begin{center}

{\Large\bf  Light-Cone Gauge for N=2 Strings~$^*$ }

\vspace{14mm}

{\Large 
Olaf Lechtenfeld~$^+$
\ \ and\ \ 
Warren Siegel }
\\[5mm]
{
C.N. Yang Institute for Theoretical Physics \\
State University of New York \\
Stony Brook, NY 11794-3840, USA 
} \\[3mm]
{Emails: olaf@lechtenfeld.de, siegel@insti.physics.sunysb.edu}
\\[5mm]

\end{center}

\vspace{2cm}

\begin{abstract}
\noindent
Covariant quantization of self-dual strings in $2{+}2$ flat dimensions reduces 
them to their zero modes, a consequence of extended world-sheet supersymmetry. 
We demonstrate how to arrive at the same result more directly by employing 
a `double' light-cone gauge. An unconventional feature of this gauge is the 
removal of anticommuting degrees of freedom by commuting symmetries and vice 
versa. The reducibility of the $N{=}4$ string and its equivalence with the 
$N{=}2$ string become apparent.
\end{abstract}

\vfill

\textwidth 6.5truein
\hrule width 5.cm
\vskip.1in

{\small
\noindent ${}^*$
Contribution to a special issue of the Russian Physics Journal \\
\noindent ${}^+$
On sabbatical leave from Institut f\"ur Theoretische Physik,
Universit\"at Hannover, Germany
}

\end{titlepage}

\noindent
{\bf 1. N=2 and N=4 strings in superconformal gauge.\ }
String theories with more than one world-sheet supersymmetry
`suffer' from the absence of higher dimensions~\cite{n2,n4}. 
Indeed, covariant quantization and naive BRST ghost-counting for the
\N2 and \NN4 superconformal algebras of constraints yield critical
dimensions of four and minus eight(!), respectively.
However, as was realized by Siegel~\cite{n4red}, the \NN4 constraints are
reducible, and the \NN4 string turns out to be the same as the \N2 string.
Yet, in contrast to the \N2 formulation, the \NN4 description is manifestly
Lorentz covariant. Since the required signature of the four-dimensional target
is $({+}{+}{-}{-})$, by `Lorentz group' we mean 
$SO(2,2)\simeq SL(2,\R)\otimes SL(2,\R)'$.
These global symmetries are not to be confused with the local R~symmetry
of \NN4 supersymmetry, denoted by~$SL(2,\R)''$.

In this letter we shall employ (Majorana) spinor notation for all Lorentz and
internal indices. We distinguish the different groups by using
\begin{equation}
\a\ \leftrightarrow\ SL(2,\R) \quad,\qquad
\ad\ \leftrightarrow\ SL(2,\R)' \quad,\qquad
\add\ \leftrightarrow\ SL(2,\R)'' \quad,\qquad
\textrm{where} \quad \a\in\{0,1\} \quad.
\end{equation}
In particular, fundamental spinors are taken to be real.\footnote{
Hence, $v^\a$ and $v^\ad$ are not related by complex conjugation,
as in $3{+}1$~dimensions.}
In the NSR formulation, both \N2 and \NN4 strings are parametrized by the 
four coordinates~$X^{\a\ad}$ plus four anticommuting NSR~fields~$\j^{\add\ad}$.
{}From the world-sheet point of view, the former are scalars while the latter
form two-component Majorana spinors, 
$\j^{\add\ad}=(\j_+^{\add\ad},\j_-^{\add\ad})$ in a Weyl basis.
With `$\pm$' we generally indicate light-cone components 
of world-sheet tensors, i.e. $\pa_\pm=\shalf(\pa_\tau\pm\pa_\sigma)$.
Sharing a world-sheet supersymmetry multiplet with~$X^{\a\ad}$, 
the anticommuting coordinates~$\j^{\add\ad}$ should also carry an undotted 
$SL(2,\R)$ index. However, its value is coupled to that of the $SL(2,\R)''$ 
index and we suppress it.
The action defines a (target-space) light-cone pairing,
\begin{equation}
(X^{0\zd},X^{1\od})\quad,\quad(\j^{\zdd\zd},\j^{\odd\od})
\qquad\textrm{and}\qquad
(X^{0\od},X^{1\zd})\quad,\quad(\j^{\zdd\od},\j^{\odd\zd}) \quad,
\end{equation}
which decomposes the variables into two $N{=}1$ light-cone sets.

Starting from the formulation with auxiliary world-sheet \N2 or
\NN4 supergravity~\cite{brink,pernici}, we advance to the superconformal gauge.
In the critical dimension, all supergravity remnants then disappear thanks to
super Weyl invariance. The residual (superconformal) freedom in this gauge 
does not prevent quantization but implies that physical states are
subject to constraints and gauge identifications. 
The \N2 constraints $(T,G^{\zdd 1},G^{\odd 0},J^{\zdd\odd})$
represent a non-degenerate subset of the (reducible) \NN4 constraints
$(T,G^{\add\a},J^{(\add\bdd)})$, 
which entails the selection of a one-parameter subgroup of the $SL(2,\R)''$ 
R-symmetry group generated by the spin-one constraints~$J^{(\add\bdd)}$. 
Because we work with real spinors it is preferable to choose a noncompact
subgroup $GL(1,\R)\subset SL(2,\R)''$.
As the R-symmetry index of~$\j^{\add\a}$ is tied to the space-time $SL(2,\R)$ 
index, the choice of~$J^{\zdd\odd}$ incidentally also breaks the Lorentz group,
\begin{equation}
SO(2,2)\ \simeq\ SL(2,\R)\otimes SL(2,\R)'\quad\longrightarrow\quad
GL(1,\R)\otimes SL(2,\R)' \quad.
\end{equation}

Being non-degenerate,
each (commuting or anticommuting) \N2 constraint essentially removes
one timelike and one spacelike degree of freedom (of matching statistics).
Hence, we expect $(T,G^{\zdd 1},G^{\odd 0},J^{\zdd\odd})$ to eliminate
{\it all\/} string coordinates~$X^{\a\ad}$ and their partners~$\j^{\add\ad}$
in the $2{+}2$ dimensional space-time. 
Indeed, this expectation is confirmed by direct analysis~\cite{bien} as well as
by amplitude computations which reveal that the \N2 string is just a point
particle encoding the dynamics of self-dual Yang-Mills and gravity~\cite{OV},
at least at tree-level.\footnote{
For a review see~\cite{marcus,dubna}. 
Quantization is detailed in~\cite{bischoff}.
The loop structure is subject of~\cite{CLN,CS}.}

Yet, it is unclear how to reproduce this result by further gauge-fixing
to a light-cone gauge.
As in other string theories, conformal reparametrizations can trivialize
one light-cone coordinate, e.g. $\pa X^{0\zd}=p^{0\zd}$, and solving
$T{=}0$ fixes a second one, e.g. $\pa X^{1\od}$. The remaining commuting 
gauge transformations, $GL(1,\R)$ generated by $J^{\zdd\odd}$, only affect 
$\j$ but not~$X$, and so cannot eliminate the pair $(X^{0\od},X^{1\zd})$. 
It seems that one is still left with 
$1{+}1$ dimensional `transverse' string excitations.
In the following, 
we resolve this contradiction by showing that the light-cone
gauge can still be used to get rid of {\it all\/} string excitations,
provided we permit {\it commuting\/} transformations to gauge-fix
{\it anticommuting\/} degrees of freedom and vice versa.
\\

\bigskip\noindent
{\bf 2. Residual N=2 superconformal symmetry.\ }
As is well known, the superconformal gauge possesses residual
gauge freedom in the form of \N2 superconformal transformations.
In light-cone world-sheet coordinates, these read
\begin{align}
\de X^{0\ad}\ &=\ (\x^+\pa_++\x^-\pa_-) X^{0\ad}\
-\ \i\eps_1^+\j_+^{\zdd\ad}\ -\ \i\eps_1^-\j_-^{\zdd\ad} \quad , \\
\de X^{1\ad}\ &=\ (\x^+\pa_++\x^-\pa_-) X^{1\ad}\
+\ \i\eps_0^+\j_+^{\odd\ad}\ +\ \i\eps_0^-\j_-^{\odd\ad} \quad , \\
\de\j_\pm^{\zdd\ad}\ &=\ \x^\pm \pa_\pm \j_\pm^{\zdd\ad}\
-\ \eps_0^\pm\pa_\pm X^{0\ad}\ -\ \l^\pm\j_\pm^{\zdd\ad} \quad , \\
\de\j_\pm^{\odd\ad}\ &=\ \x^\pm \pa_\pm \j_\pm^{\odd\ad}\
+\ \eps_1^\pm\pa_\pm X^{1\ad}\ +\ \l^\pm\j_\pm^{\odd\ad} \quad ,
\end{align}
where the commuting parameter functions $(\x^+,\l^+)$ and $(\x^-,\l^-)$
depend only on $\t{+}\s$ and $\t{-}\s$, respectively, while the anticommuting
parameter functions $(\eps^+_0,\eps^+_1)$ and $(\eps^-_0,\eps^-_1)$
do likewise.

The transformations parametrized by
$(\x^+,\eps^+_0,\eps^+_1,\l^+)$ are generated by the left-moving \N2 currents
$(T,G^{\odd 0},G^{\zdd 1},J^{\zdd\odd})$, in that order. 
The other half (carrying a `$-$' superscript) goes with a right-moving 
copy of those currents. We shall focus on the left-movers and drop all
world-sheet indices. Then the \N2 generators read
\begin{align}
T\ &=\ \pa X^{0\zd} \pa X^{1\od} - \pa X^{0\od} \pa X^{1\zd} +
       \i\j^{\zdd\zd} \pa\j^{\odd\od} - \i\j^{\zdd\od} \pa\j^{\odd\zd} -
       \i\j^{\odd\zd} \pa\j^{\zdd\od} + \i\j^{\odd\od} \pa\j^{\zdd\zd} \quad,
\label{T} \\
G^{\zdd 1}\ &=\ \j^{\zdd\zd} \pa X^{1\od} - \j^{\zdd\od} \pa X^{1\zd} \quad,
\label{G01} \\
G^{\odd 0}\ &=\ \j^{\odd\zd} \pa X^{0\od} - \j^{\odd\od} \pa X^{0\zd} \quad,
\label{G10} \\
J^{\zdd\odd}\ &=\ \j^{\zdd\zd} \j^{\odd\od} - \j^{\zdd\od} \j^{\odd\zd} \quad.
\label{J01}
\end{align}
As usual, $T$ is associated with conformal coordinate transformations
while $J^{\zdd\odd}$ holomorphically rescales the NSR fields only.
Note that the $GL(1,\R)$ current is antihermitian.
More interesting is the action of the anticommuting generators, 
depicted in full detail as
\begin{equation} \label{cd}
\begin{CD}
X^{0\zd} @>{G^{\zdd 1}}>> \j^{\zdd\zd} @>{\cdot\;G^{\odd 0}}>> \pa X^{0\zd} \\
@V{\Big\uparrow}VV @V{\Big\uparrow}VV @V{\Big\uparrow}VV \\
X^{1\od} @>{G^{\odd 0}\cdot}>> \j^{\odd\od} @>{G^{\zdd 1}}>> \pa X^{1\od}
\end{CD}
\qquad\textrm{and}\qquad
\begin{CD}
X^{0\od} @>{\cdot\;G^{\zdd 1}}>> \j^{\zdd\od} @>{G^{\odd 0}}>> \pa X^{0\od} \\
@V{\Big\uparrow}VV @V{\Big\uparrow}VV @V{\Big\uparrow}VV \\
X^{1\zd} @>{G^{\odd 0}}>> \j^{\odd\zd} @>{G^{\zdd 1}\cdot}>> \pa X^{1\zd}
\end{CD} 
\end{equation}
where the vertical arrows relate (target-space) light-cone conjugate\footnote{
Two variables are light-cone conjugate when their part of the (target-space)
metric is $\left(\begin{smallmatrix} 0 & 1 \\ 1 & 0 \end{smallmatrix}\right)$.
It implies that the momentum and annihilation parts of one variable are 
canonically conjugate to the position and creation parts of the other.}
variables.
\\

\bigskip
\noindent
{\bf 3. Gauge fixing part one -- the conventional part.\ }
As is well-known, the residual-invariance generators play a double role
because their vanishing has to be imposed as a constraint on the theory.
In the bosonic string, for example, one employs $T$ first to gauge-fix
a light-cone coordinate, say~$\pa X^+=p^+$, and then a second time to
solve $T{=}0$ for the light-cone conjugate coordinate~$\pa X^-$, in effect
getting rid of $1{+}1$~dimensions.
This mechanism generalizes to the $N{=}1$ supersymmetric case,
where an anticommuting generator~$G$ allows one to transform $\j^+$ to zero
and obtain $\j^-$ as a function of the transversal coordinates by solving
$G{=}0$.

Let us do the same for the light-cone pair
$(X^{0\zd},X^{1\od})$ plus $(\j^{\zdd\zd},\j^{\odd\od})$
by making use of the generators $T$ and~$G^{\odd 0}$.
By additional dots in the left part of~(\ref{cd}) we indicate
the function of the anticommuting generator.
In this way we arrive at
\begin{align} \label{slice1}
\pa X^{0\zd}\ &=\ p^{0\zd} \quad,\qquad
\pa X^{1\od}\  =\ \pinv (\pa X^{0\od} \pa X^{1\zd}
	+ \i\j^{\zdd\od} \pa\j^{\odd\zd} + \i\j^{\odd\zd} \pa\j^{\zdd\od}) \\
\label{slice2}
\j^{\zdd\zd}\ &=\ 0 \quad,\qquad\quad
\j^{\odd\od}\  =\ \pinv\;\j^{\odd\zd} \pa X^{0\od} \quad,
\end{align}
and the remaining \N2 generators simplify to
\begin{equation} \label{constraints}
G^{\zdd 1}\ =\ -\j^{\zdd\od} \pa X^{1\zd} 
\qquad\textrm{and}\qquad
J^{\zdd\odd}\ =\ -\j^{\zdd\od} \j^{\odd\zd} \quad.
\end{equation}
Although this reasoning is purely classical, it can be incorporated in the
quantum theory by reading~(\ref{slice1})--(\ref{constraints}) as operator
statements and replacing $p^{0\zd}$ by the c-number $k^{0\zd}$.
\\

\bigskip\noindent
{\bf 4. Chiral bosonization.\ }
In order to get rid of the remaining pair
$(X^{0\od},X^{1\zd})$ plus $(\j^{\zdd\od},\j^{\odd\zd})$
we must attempt to employ the {\it commuting\/} generator~$J^{\zdd\odd}$
to eliminate the {\it anticommuting\/}~$\j$'s and then kill the $X$'s 
with~$G^{\zdd 1}$.
On the classical level this makes no sense because ordinary and Grassmann
numbers are not related in any way. Upon quantization, however, this
distinction blurs: fermionic creation and annihilation operators may be
represented by finite matrices (with commuting entries), simply leading to
multi-component wave functions. In two dimensions, one even has a direct
relation between commuting and anticommuting fields by means of bosonization.
Following this idea, we describe the $(\j^{\zdd\od},\j^{\odd\zd})$ system
by a chiral boson~$\phi$ (normal-ordering implied),
\begin{equation} \label{bosonize}
\j^{\zdd\od}\ =\ \e^{-\phi} 
\qquad\textrm{and}\qquad
\j^{\odd\zd}\ =\ \e^{+\phi} \quad,
\qquad\textrm{so that}\qquad
\j^{\odd\zd}\,\j^{\zdd\od}\ =\ \i\,\pa\phi \quad.
\end{equation}
Note that $\phi$ is not an angle but a scale; there are no winding modes.
Let us remark on the statistics. 
For more than one pair of anticommuting fields the above equations have
to be supplemented by suitable cocycle (a.k.a. Jordan-Wigner or Klein) factors
which ensure that the different pairs mutually anticommute.
Since in the present case, however, only a single pair of NSR~fields is left,
its statistics is irrelevant and we can fully describe it by the 
chiral commuting~$\phi$.

Invoking the decomposition
\begin{equation}
\phi\ =\ \phi_< + q + (\t{+}\s)p + \phi_> 
\end{equation}
into negative-, zero-, and positive-frequency parts,\footnote{
Positive frequency means positive Fourier modes, i.e. the annihilation part.
Zero modes obey $[q,p]=\i$.}
the exponential operators read
\begin{equation}
\e^{n\phi}\ =\ \e^{n\phi_<}\;\e^{nq+n(\t{+}\s)p}\;\e^{n\phi_>}\ =\
\e^{n\phi_<}\;\e^{\frac{n}{2}q}\;\e^{n(\t{+}\s)p}\;\e^{\frac{n}{2}q}\;
\e^{n\phi_>} \quad.
\end{equation}
The $\phi$ Fock space is generated by $\phi$ creation operators 
(contained in~$\phi_<$) acting on a momentum eigenstate~$|a\rangle$, defined by
\begin{equation}
p\,|a\rangle\ =\ a\,|a\rangle 
\end{equation}
and created from the vacuum state via
\begin{equation}
|a\rangle\ =\ \e^{\i a q}\,|0\rangle\ =\ 
\e^{\i a\phi}(\t{\to}{-}\infty)\,|0\rangle \quad.
\end{equation}
On such a state, exponential operators act as
\begin{equation}
\e^{n\phi}\,|a\rangle\ =\ 
\e^{n(a-\i\frac{n}{2})(\t+\s)}\;\e^{n\phi_<}\, |a{-}n\i\rangle \quad.
\end{equation}

{}From this we learn two things.
First, $\i p$ is nothing but fermion number because
the NSR~fields $\e^{\pm\phi}$ shift the eigenvalue by~$\pm1$ unit.~\footnote{
Due to the indefinite target-space metric there is no conflict with the
anti-hermiticity of~$\i p$.}
Second, $a$ is tied to the monodromy for the NSR~fields,
\begin{equation}
\e^{\pm\phi}(\s{+}2\pi)\,|a\rangle\ =\ 
-\,\e^{\pm 2\pi a}\,\e^{\pm\phi}(\s)\,|a\rangle \quad.
\end{equation}
It should be noted that, due to the reality of the NSR~fields, the monodromy
group is~$\R_+$ and not $U(1)$. The Fourier modes are
\begin{equation}
\j^{\zdd\od}_{m+\frac12+\i a}
\qquad\textrm{and}\qquad
\j^{\odd\zd}_{m+\frac12-\i a}
\end{equation}
with $m\in\Z$, so all monodromy sectors are NS-like. 
Due to the spectral flow isometry, which acts as
\begin{equation}
\j\ \longmapsto\ \e^{-\i a q}\,\j\,\e^{\i a q}\ =\ \e^{\pm(\t+\s)a}\,\j \quad,
\end{equation}
all monodromy sectors are equivalent. In the following, we choose $a{=}0$.
\\

\bigskip\noindent
{\bf 5. Gauge fixing part two -- the unconventional part.\ }
How is the gauge-fixing accomplished in this framework?
As for example in the Gupta-Bleuler method, we shall impose the remaining
constraints 
\begin{equation}
G^{\zdd 1}\ =\ -\e^{-\phi}\,\pa X^{1\zd}
\qquad\textrm{and}\qquad
J^{\zdd\odd}\ =\ \i\,\pa\phi 
\end{equation}
not as operator equations but rather demand that their 
positive- and zero-frequency parts annihilate the physical states,\footnote{
except for the zero mode $J_0^{\zdd\odd}=\i p$ as seen above.
$G^{\zdd 1}$ has no zero modes.}
\begin{equation}
G_\ge^{\zdd 1}(\s)\,|\textrm{phys}\rangle\ =\ 0
\qquad\textrm{and}\qquad
J_\ge^{\zdd\odd}(\s)\,|\textrm{phys}\rangle\ =\ \i a\,|\textrm{phys}\rangle 
\quad.
\end{equation}
If a constraint is not self-conjugate its conjugate will create a gauge 
invariance which allows us to further restrict the physical states.
Our state space consists of the zero modes and excitations of the free fields
$\phi$, $X^{0\od}$, and~$X^{1\zd}$. In other words, its basis is generated
by acting with their creation operators on the momentum eigenstates
$|a,k^{0\od},k^{1\zd}\rangle$. 

We first attend to the commuting generator, $J^{\zdd\odd}=\i\pa\phi$.
Classically, it effects local translations of~$\phi$.
On quantum states, we demand
\begin{equation}
\pa\phi_>\,|\textrm{phys}\rangle\ =\ 0 \quad,
\end{equation}
which removes all $\phi$ creation operators from $|\textrm{phys}\rangle$.
The $\phi$ content of $|\textrm{phys}\rangle$ is thus reduced to its zero mode,
whose value we chose to be $a{=}0$.

Finally, we expose the action of the anticommuting generator, 
$G^{\zdd 1}=-\e^{-\phi}\pa X^{1\zd}$. Since 
\begin{equation} \label{nozero}
\e^{-\phi}\,|0\rangle\ =\ 
\e^{-\frac{\i}{2}(\t+\s)}\;\e^{-\phi_<}\,|\i\rangle
\end{equation}
the r.h.s. contains only negative-frequency (and no zero-mode) parts.
Therefore,
$G_\ge^{\zdd 1}$ on physical states will involve all positive modes
of~$\pa X^{1\zd}$ but no others. Consequently, the requirement
\begin{equation}
\bigl( \pa X^{1\zd}\,\e^{-\phi} \bigr)_\ge\,|\textrm{phys}\rangle\ =\ 0
\end{equation}
eliminates all $X^{0\od}$ creation operators from $|\textrm{phys}\rangle$.
By conjugation, the associated gauge symmetry allows us to gauge away all
$X^{1\zd}$ creation operators as well.
We are now left with 
\begin{equation} \label{phys}
|\textrm{phys}\rangle\ =\ |0,k^{0\od},k^{1\zd}\rangle \quad.
\end{equation}

Finally, we may include the momenta $k^{0\zd}$ and~$k^{1\od}$ back into
the parametrization of the physical states.
Then, the `$11$' coordinates from
(\ref{slice1}) and~(\ref{slice2}) may be expressed as\footnote{
The classical constraint $T{=}0$ gets modified to $T{=}a^2$.
This cancels the contribution of the NSR~momenta to~$L_0$.}
\begin{equation}
\pa X^{1\od}\ =\ \kinv ( \pa X^{0\od} \pa X^{1\zd} + a^2 - \pa\phi\pa\phi )
\qquad\textrm{and}\qquad
\j^{\odd\od}\ =\ \kinv\;\pa X^{0\od}\,\e^{+\phi} \quad.
\end{equation}
On the states~(\ref{phys}) only the non-positive modes contribute.
The zero mode of $\pa X^{1\od}$ yields the mass-shell condition
$k^{0\zd}k^{1\od}-k^{0\od}k^{1\zd}=0$.

Summarizing, the `double' light-cone gauge reduces the \N2 string
degrees of freedom to the zero modes of its bosonic coordinates
and puts them on mass-shell. As in the covariant treatment, 
a massless free boson makes up the entire spectrum of states.

We finally remark that the elimination mechanism works differently for bosons 
and fermions. It is crucial that the bosons come in light-cone pairs, i.e.
they support an indefinite space-time metric, while each (light-cone conjugate)
fermion pair gives rise (via bosonization) to a single self-conjugate boson,
so that these chiral bosons support only a Euclidean metric.
For this reason it is impossible\footnote{
It is possible only for the zero modes.} 
to employ a commuting gauge invariance to gauge-fix only half 
of a fermionic pair and eliminate the other half through the constraint, 
as is done for the $X$'s.
After both gauge-fixing and imposing the constraints, however, the counting
for bosons and fermions again coincides.
\\

\bigskip
\noindent
{\bf 6. The N=4 case.\  }
To streamline the equations for the \NN4 string, 
we only display the left-moving degrees of freedom 
and drop the world-sheet $(\pm)$ indices. 
The right-moving part behaves completely analogously.
The \NN4 superconformal transformations 
\begin{align} \label{N41}
\de X^{0\ad}\ &=\ \x\,\pa X^{0\ad}\
+\ \i\eps_1^\zdd\j^{\odd\ad}\ -\ \i\eps_1^\odd\j^{\zdd\ad} \quad , \\
\de X^{1\ad}\ &=\ \x\,\pa X^{1\ad}\
+\ \i\eps_0^\zdd\j^{\odd\ad}\ -\ \i\eps_0^\odd\j^{\zdd\ad} \quad , \\
\de\j^{\zdd\ad}\ &=\ \x\,\pa \j^{\zdd\ad}\
+\ \eps_0^\zdd \pa X^{0\ad}\ +\ \eps_1^\zdd \pa X^{1\ad}\
+\ \l^{\zdd\zdd} \j^{\odd\ad}\ -\ \l^{\zdd\odd} \j^{\zdd\ad} \quad , \\
\de\j^{\odd\ad}\ &=\ \x\,\pa \j^{\odd\ad}\
+\ \eps_0^\odd \pa X^{0\ad}\ +\ \eps_1^\odd \pa X^{1\ad}\
+\ \l^{\odd\zdd} \j^{\odd\ad}\ -\ \l^{\odd\odd} \j^{\zdd\ad} 
\label{N44}
\end{align}
involve the four commuting left-moving parameters $(\x,\l^{(\add\bdd)})$
and four anticommuting left-moving parameters $(\eps_\a^\add)$.
Clearly, the relation with the \N2 parameters is
\begin{equation}
\x=\x \quad,\qquad
\l=\l^{\zdd\odd}=\l^{\odd\zdd} \quad,\qquad
\eps_0= \eps_0^\zdd \quad,\qquad
\eps_1= \eps_1^\odd \quad.
\end{equation}
The \NN4 constraints which generate the above transformations
consist of the \N2 set
(\ref{T})--(\ref{J01}) enlarged by
\begin{align}
\eps_0^\odd:\
G^{\zdd 0}\ &=\ \j^{\zdd\zd} \pa X^{0\od}\ -\ \j^{\zdd\od} \pa X^{0\zd} \quad,
\label{G00} \\
\eps_1^\zdd:\
G^{\odd 1}\ &=\ \j^{\odd\zd} \pa X^{1\od}\ -\ \j^{\odd\od} \pa X^{1\zd} \quad, 
\label{G11} \\
\l^{\odd\odd}:\
J^{\zdd\zdd}\ &=\ 2\,\j^{\zdd\zd} \j^{\zdd\od} \quad,
\label{J00} \\
\l^{\zdd\zdd}:\
J^{\odd\odd}\ &=\ 2\,\j^{\odd\zd} \j^{\odd\od} \quad.
\label{J11}
\end{align}

Let us investigate if the non-negative-frequency parts of the additional 
four constraints imply any further restrictions on the physical states 
of the \N2 string. On the gauge slice (\ref{slice1}) and (\ref{slice2})
the above generators reduce to
\begin{align}
G^{\zdd 0}\ &=\ -\e^{-\phi}\,k^{0\zd} \quad, \\
G^{\odd 1}\ &=\ -\e^{+\phi}\,(\pa\phi\pa\phi -a^2)/k^{0\zd} \quad, \\
J^{\zdd\zdd}\ &=\ 0 \quad, \\
J^{\odd\odd}\ &=\ 2\,:\e^{+\phi} \e^{+\phi}:\,\pa X^{0\od}/k^{0\zd}\ =\ 0\quad.
\end{align}
On the physical states~$|0,k^{\a\ad}\rangle$,
only negative-frequency modes will be created,
and $G^{\add\b}$ do not contain zero modes.
Therefore, indeed
\begin{equation}
G_\ge^{\zdd 0}\,|0,k^{\a\ad}\rangle\ =\ 0\ =\
G_\ge^{\odd 1}\,|0,k^{\a\ad}\rangle \quad.
\end{equation}
Hence, the additional \NN4 constraints are obsolete on the physical
states of the \N2 string, as expected.
\\

\bigskip\noindent
{\bf Acknowledgements.\  }
O.L. thanks Gordon Chalmers for collaboration at an early stage.
This work is partially supported by DFG grant Le 838/7-1, by a sabbatical
research grant of the Volkswagen-Stiftung, and by NSF grant PHY-0098527.

\vfill\eject

\end{document}